\documentclass[a4paper,11pt]{article}
\pdfoutput=1 % if your are submitting a pdflatex (i.e. if you have
\usepackage{jcappub} % for details on the use of the package, please
\usepackage[utf8]{inputenc}
\usepackage[english]{babel}
 
\usepackage[T1]{fontenc} % if needed

\usepackage{hyperref}

\usepackage{cmap}

\title{\boldmath Cosmology based on $f(R)$ gravity
with ${\cal O}(1)$ eV sterile neutrino}

\author[a,b,1]{Anton  S. Chudaykin,\note{Corresponding author.}}
\author[a,b]{Dmitry S. Gorbunov,}
\author[c,d]{Alexei A. Starobinsky,}
\author[e,b]{Rodion A. Burenin}

\affiliation[a]{Institute for Nuclear Research of the Russian Academy of Sciences,\\60th October Anniversary prospect 7a, Moscow 117312, Russian Federation}
\affiliation[b]{Moscow Institute of Physics and Technology,\\Institutsky per. 9, Dolgoprudny 141700, Russian Federation}
\affiliation[c]{L. D. Landau Institute for Theoretical Physics of the Russian Academy of Sciences, \\ Moscow 119334, Russian Federation}
\affiliation[d]{Kazan Federal University, \\ Kazan 420008, Republic of Tatarstan, Russian Federation}
\affiliation[e]{Space Research Institute of the Russian Academy of
  Sciences (IKI),\\ Moscow, ul.~Profsoyuznaya, 84/32, 117997, Russian Federation}

\emailAdd{chudy@ms2.inr.ac.ru}
\emailAdd{gorby@ms2.inr.ac.ru}
\emailAdd{alstar@landau.ac.ru}
\emailAdd{rodion@hea.iki.rssi.ru}

\abstract{We address the cosmological role of an additional  ${\cal O}(1)$ eV
  sterile neutrino in modified gravity models. We confront the present
  cosmological data with predictions of the FLRW cosmological model
  based on a variant of $f(R)$ modified gravity proposed by one of the
  authors previously. This viable cosmological model which deviation
  from general relativity with a cosmological constant $\Lambda$
  decreases as $R^{-2n}$ for large, but not too large values of the
  Ricci scalar $R$ (while no $\Lambda$ is introduced by hand at small
  $R$) provides an alternative explanation of present dark energy and the
  accelerated expansion of the Universe (the case $n=2$ is considered in
  the paper). Various up-to-date cosmological data sets exploited include
  measurements of the cosmic microwave background (CMB) anisotropy, the
  CMB lensing potential, the baryon acoustic oscillations (BAO), the
  cluster mass function and the Hubble constant. We find that the CMB+BAO
  constraints strongly restrict the sum of neutrino masses from above. This
  excludes values of the model parameter $\lambda\sim 1$ for which distinctive
  cosmological features of the model are mostly pronounced as compared to
  the $\Lambda$CDM model, since then free streaming damping of perturbations
  due to neutrino rest masses is not sufficient to compensate their
  extra growth occurring in $f(R)$ modified gravity. Thus, in the gravity sector
  we obtain $\lambda>8.2$ ($2\sigma$) with the account of systematic
  uncertainties in galaxy cluster mass function measurements and
  $\lambda>9.4$ ($2\sigma$) without them. At the same time in the latter
  case we find for the sterile neutrino mass
  $0.47\,\,\rm{eV}$$\,<\,$$m_{\nu,\,\rm{sterile}}$$\,<\,$$1\,\,\rm{eV}$
  ($2\sigma$) assuming that the sterile neutrinos are thermalized and the
  active neutrinos are massless, not significantly larger than in the
  standard $\Lambda$CDM with the same data set:
  $0.45\,\,\rm{eV}$$\,<\,$$m_{\nu,\,\rm{sterile}}$$\,<\,$$0.92\,\,\rm{eV}$
  ($2\sigma$). However, a possible discovery of a sterile neutrino with the mass
  $m_{\nu,\,\rm{sterile}} \approx 1.5$\,eV motivated by various anomalies in
  neutrino oscillation experiments would favor cosmology based on $f(R)$ gravity
  rather than the $\Lambda$CDM model.

}

\begin{document}
\maketitle
\flushbottom

\section{Introduction}
\label{sec:intro}

The fact that the present Universe is undergoing an accelerated expansion
is firmly established by numerous observational data. The standard
$\Lambda$-Cold-Dark-Matter ($\Lambda$CDM) cosmological model can
explain the cosmic acceleration at expense of introducing a new
fundamental physical parameter, the cosmological constant
$\Lambda$. Its observed value is much smaller than any other energy
scale of the fundamental physical interactions, that presents a great
challenge for the theoretical elementary particle physics.
%Clearly, theoretical physics is in desperate strait.

A stage similar to the present accelerated expansion one, dubbed
inflation, is believed to happen in the very early Universe.  We know
that the source of inflation may not be identical to the cosmological
constant since the inflaton field -- primordial Dark Energy (DE) -- was
evolving and unstable. This qualitative analogy provides with an
additional argument in favor of non-stationary models of the present
DE alternative to $\Lambda$.

In this paper, we consider so-called $f(R)$ gravity (see
e.g.~\cite{fullaboutfR,SF2010,FT2010,NO2011} for reviews,
and~\cite{chameleonmech,stabconditions,Starmodel} for the first viable
cosmological models relevant for the present Universe), which modifies General
Relativity (GR) by replacing the scalar curvature (the Ricci scalar) $R$
with a new phenomenological function $f(R)$ in the Einstein-Hilbert
action. It represents a special case of more general scalar-tensor
Brans-Dicke theory~\cite{BGtheory} with the Brans-Dicke parameter
$\omega_{BD}=0$~\cite{wBG=0}. Cosmological models based of this
modified gravity can explain the present cosmic acceleration without
introducing $\Lambda$, so we can put $f(0)=0$. There is a new scalar
degree of freedom in the gravity sector dubbed {\it
  scalaron}~\cite{Starinflation} responsible for extra growth of
matter density perturbations in $f(R)$ models. That is the most
dramatic difference from the $\Lambda$CDM model which we have to cope
with.

For an $f(R)$ model to be phenomenologically viable and theoretically
consistent and to solve the above difficulties, it
should satisfy a list of viability conditions. First, an $f(R)$ should
satisfy the necessary  conditions in the region of relevant values of $R$:
\begin{equation}  \label{eq:math:ex3}
f'(R)>0\,,
\qquad
f''(R)>0\,,
\end{equation}
hereinafter prime denotes a derivative with respect to argument
$R$. The first condition in \eqref{eq:math:ex3} means that the gravity is
attractive and graviton is not a ghost. The second condition guarantees
that scalaron is not a tachyon both in the Minkowski space-time and in the regime
of small deviations from GR. Note that it is necessary to keep conditions
\eqref{eq:math:ex3} for all values of $R$ during the matter- and radiation-dominated
stages in order to avoid the Dolgov--Kawasaki instability~\cite{Dolgovinst}. If
one wants to incorporate the early-time inflation, the range of $R$
has to be extended accordingly, see discussion in Sec.~\ref{sec:fon}.

Second, the existence of the new additional degree of freedom imposes a
number of special conditions on the functional form of $f(R)$ for $R\gg
R_{0}$~\cite{Starmodel}:
\begin{equation}  \label{eq:math:ex4}
|f(R)-R|\ll R\,,
\qquad
|f'(R)-1|\ll1\,,
\qquad
f''(R)R\ll1\,,
\end{equation}
where $R_{0}$ is the present Ricci curvature. These conditions
guarantee the correct Newtonian limit for the matter-dominated stage in
the past and smallness of non-GR corrections to a space-time
background metric for a more general background of compact astrophysical
objects in the present Universe. The third condition in
\eqref{eq:math:ex4} implies that the Compton wavelength of the
scalaron field is much less than the curvature radius of the
background metric. It ensures the absence of extra growth of the matter
perturbations in high-density regions that is necessary to satisfy local gravity constraints (LGC). On the other hand, in principle,  this condition may be violated at $R\sim R_0$ due to dependence of the effective scalaron mass on $R$
which, in turn, is determined by the matter density in the regime of small deviations
from GR. In cosmology, such effect is often called the chameleon mechanism~\cite{chameleonmech}, though it occurs in many other areas of physics, too, c.f. the well-known dependence of the plasmon mass on density in plasma physics. It is important for understanding of behavior of the cosmological perturbations in $f(R)$ gravity (see discussion in Sec.~\ref{sec:fluct}).

If the above constraints are satisfied, cosmological models based on $f(R)$ gravity
can describe FLRW background expansion history similarly to that of the $\Lambda$CDM model. However, inhomogeneous metric fluctuations evolve differently. In particular, matter density perturbations grow faster on scales smaller than the Compton wavelength of the scalaron field that occurs at recent redshifts. One needs something to compensate for this extra
growth. For instance, neutrino rest masses can do this job. The free streaming
of the massive neutrinos suppresses the structure formation on small
scales. Hence, if one adjusts the neutrino masses in the $f(R)$ gravity, the net
result can be zero, because the $f(R)$ modification and neutrino masses
play opposite roles in the evolution of matter density perturbations
on small scales~\cite{massneutrino+f(R)}.

The same mechanism works for a {\cal O}(1)\,eV sterile neutrino added to the Standard Model of elementary particles. If mixing with the active neutrino is
not extremely small, the sterile neutrinos are produced in the primordial
plasma and get thermalized in the early Universe before the active
neutrino decoupling. While relativistic they contribute to the
radiation component as one additional neutrino species. In particular,
this component increases the Universe expansion rate at the Big Bang
Nucleosynthesis (BBN) epoch. The recent reanalysis of the primordial helium
abundance permits the existence of one extra neutrino species \cite{BBN}: an
effective number of neutrinos is $N_{eff}=3.58\pm0.40 (2\sigma)$ for the
neutron lifetime $\tau_{n}=880.1\pm1.1$s, while the standard three
active neutrinos give $N_{eff}=3.046$~\cite{SBBN}.

The light sterile neutrinos are interesting because of anomalous
results obtained by several neutrino oscillation experiments which do
not fit to the three neutrinos oscillation pattern and ask for one (or
two) more light neutrinos\,\cite{Agashe:2014kda}.
In particular, the so-called gallium anomaly observed by
GALLEX\,\cite{GALLEX1,GALLEX2} and SAGE\,\cite{SAGE1,SAGE2} experiments
is nicely explained as the electron neutrino oscillations
into sterile neutrino of $1.5$\,eV mass\,\cite{Giunti:2010zu}.
The reactor antineutrino anomaly (disappearance of electron
antineutrinos from nuclear reactors) \cite{reactor1,reactor2}
is consistent with sterile neutrino of the same mass, while account
for other anomalies from accelerator experiments shifts the mass in the
combined fit to $1-1.3$\,eV, see \cite{Agashe:2014kda} for details.

In this paper we confront the most recent observational data with predictions of
cosmology based on $f(R)$ gravity and supplemented with light sterile neutrinos. In Sec.~\ref{sec:fon} we present the $f(R)$ cosmological model and describe the
cosmological background (homogeneous Universe) evolution. In
Sec.\,~\ref{sec:fluct} we consider matter density perturbations. We fit
the model predictions to the observational data in
Sec.\,\ref{sec:cosmo}.  For the $f(R)$ model we outline the allowed region in the model parameter space using modified MGCAMB and
CosmoMC.  We find that $1.5$\,eV sterile neutrino is better consistent
with DE models based on the $f(R)$ gravity rather than with the standard $\Lambda$CDM.
Among all the cosmological data used, the most important appear to be those from
observations of galaxy clusters which trace the evolution of density perturbations. In contrast
to the paper~\cite{Star_1eV} where the same problem was studied, we do not use
the power spectrum of matter density perturbations obtained from galaxy clustering data to avoid
problems with the bias parameter. Instead, we employ more recent and
accurate results on the abundance of galaxy clusters.

%%%%%%%%%%%%%%%%%%%%%%%%%%%%%%%%%%%%%%%%%%%%%%%%%%%%%%%%%%%%%%%%%%%%%%%%
\section{Background Universe}
\label{sec:fon}

We define $f(R)$ gravity by the following action
\begin{equation}  \label{eq:math:ex1}
S=\frac{1}{2\kappa^{2}}\int d^{4}x\sqrt{-g}f(R)+S_{m}\,,
\end{equation}
where $\kappa^{2}/(8\,\pi)\equiv G$ is the Newton gravitational constant
and $S_{m}$ is the action of matter fields all minimally
coupled to gravity.

We take the $f(R)$ model~\cite{Starmodel}
\begin{equation}  \label{eq:math:ex5}
f(R)=R+\lambda
R_{s}\left[\left(1+\frac{R^2}{R_{s}^2}\right)^{-n}-1\right]\,,
\end{equation}
where $n, \lambda, R_{s}$ are model parameters. Strictly speaking, the
model \eqref{eq:math:ex5} has to be modified at very large values of
$R$, e.g. supplemented with the term $R^{2}/6M^{2}$ borrowed from the
inflationary model~\cite{Starinflation} where $M$ is the inflaton
mass. This term solves problems discussed
in Ref.~\cite{curingsingular}: the scalaron mass exceeding the Planck mass
and a weak curvature singularity at some finite time in the past. The
value of $M$ should be sufficiently large in order to pass laboratory
and Solar system tests of gravity, namely one has $M>10^{-2.5}$ eV according to
Cavendish-type experiment~\cite{Cavendish}. However, any type of inflation in the early Universe (driven by either scalaron
or another field) imposes a much higher upper limit on $M$, only
several orders of magnitude less than the Planck mass. In particular,
in the latter case, one has $M\gg H_{inf}$ where $H_{inf}$ is the Hubble
parameter at the end of inflation. As a result, this high-$R$
correction to $f(R)$ becomes negligible for the low-$R$ cosmology we
are interested in. Also, the
function \eqref{eq:math:ex5} should be modified and conditions
\eqref{eq:math:ex3} should be revised for $R<R_{0}$ including the
region $R<0$ ($R$ becomes negative during
post-inflationary evolution, see~\cite{curingsingular} for detailed
study of this issue). Once more, this change does not affect our case
where $R\ge R_0$.

%Our "unified" model \eqref{eq:math:ex2} describes both inflation in the early Universe and present DE driving recent acceleration of the Universe and leads to %slightly different predictions for the primordial perturbation spectra, as compared to the purely inflationary model ($\lambda R_{s}=0$). Furthermore, in this %unified model the first term in \eqref{eq:math:ex2} should be modified for $R<R_{0}$. Therefore, conditions \eqref{eq:math:ex3} should be reconsidered and %changed for $R<R_{0}$ including the region $R<0$. It is occurred because $R$ becomes negative during post-inflationary reheating. For more detailed %information about this topic see \cite{curingsingular}.\par

%Because of we are interested in the dynamics of the late time Universe we can neglect the $R^{2}$ term and use \eqref{eq:math:ex2} without last term in future:
%\begin{equation}  \label{eq:math:ex5}
%f(R)=R+\lambda R_{s}\Big{[}\Big{(}1+\frac{R^2}{R_{s}^2}\Big{)}^{-n}-1\Big{]}.
%\end{equation} \par

%%%%%%%%%%%%%%%%%%%%%%%%%%%%%%%%%%%%%%%%%%%%%%%%%%%%%%%%%%%%%
\subsection{Field equations}

We derive field equations by varying the action \eqref{eq:math:ex1}
with respect to space-time metric\footnote{We use the sign conventions in which the metric given by
  $ds^{2}=-dt^{2}+a^{2}(t)d\vec{x}^{2}$, where $a(t)$ is the scale factor.} $g_{\mu\nu}$
%\begin{equation}  \label{eq:math:ex6}
%R_{\mu\nu}-\frac{1}{2}Rg_{\mu\nu}=\frac{1}{f'(R)}\Big[\frac{1}{2}g_{\mu\nu}(f(R)-Rf'(R))+(\nabla_{\mu}\nabla_{\nu}-g_{\mu\nu}\Box)f'(R)\Big]+\frac{\kappa^{2}}{f'(R)}T_{\mu\nu}^{matter}
%\end{equation} \par
\begin{equation}  \label{eq:math:ex6}
f'R_{\mu\nu}-\frac{1}{2}fg_{\mu\nu}+(g_{\mu\nu}\Box-\nabla_{\mu}\nabla_{\nu})f'=\kappa^{2}T_{\mu\nu}^{(M)}\,,
\end{equation}
where $R_{\mu\nu}$ is the Ricci tensor, $\nabla_{\mu}$ is
the covariant derivative  associated with  the metric $g_{\mu\nu}$,
$\Box\phi\equiv g^{\mu\nu}\nabla_{\mu}\nabla_{\nu}\phi$ and
$T_{\mu\nu}^{(M)}$ is the matter stress-energy tensor.

We obtain two gravitational field equations from diagonal elements of
\eqref{eq:math:ex6}:
\begin{equation}  \label{eq:math:ex9}
3H^{2}f'-\frac{1}{2}(Rf'-f)+3H\dot{f'}=\kappa^{2}\rho_{m}\,,
\end{equation}
\begin{equation}  \label{eq:math:ex10}
(\dot{H}+3H^{2})f'-\frac{1}{2}f-\ddot{f'}-2H\dot{f'}=\kappa^{2}P_{m}\,,
\end{equation}
where $H=\dot{a}/a$ is the Hubble parameter,  $\rho_{m}$ and
$P_{m}$ are the energy density and pressure of non-relativistic matter
($P_{m}=0$), respectively, and hereafter dot denotes
derivative with respect to the cosmic time $t$.

We can rewrite \eqref{eq:math:ex6} in the following Einsteinian form,
\begin{equation}  \label{eq:math:ex7}
R_{\mu\nu}-\frac{1}{2}g_{\mu\nu}R=\kappa^{2}\left(
T_{\mu\nu}^{(M)}+T_{\mu\nu}^{(DE)}\right)\,,
\end{equation}
where
\begin{equation}  \label{eq:math:ex8}
\kappa^{2}T_{\mu\nu}^{(DE)}\equiv\frac{1}{2}g_{\mu\nu}(f-R)-(g_{\mu\nu}\Box-\nabla_{\mu}\nabla_{\nu})f'+R_{\mu\nu}(1-f')\,.
\end{equation}
Then from \eqref{eq:math:ex9}, \eqref{eq:math:ex10} and
\eqref{eq:math:ex8} we obtain the effective DE density
$\rho_{DE}$  and  pressure $P_{DE}$:
\begin{equation}  \label{eq:math:ex11}
\kappa^{2}\rho_{DE}=-3H\dot{f'}+3(H^{2}+\dot{H})(f'-1)-\frac{1}{2}(f-R)\,,
\end{equation}
\begin{equation}  \label{eq:math:ex12}
\kappa^{2}P_{DE}=\ddot{f'}(R)+2H\dot{f'}-(3H^{2}+\dot{H})(f'-1)+\frac{1}{2}(f-R)\,.
\end{equation}
Then we define the equation-of-state parameter $\omega_{DE}$ for the DE component by
\begin{equation}  \label{eq:math:ex13}
\omega_{DE}\equiv\frac{P_{DE}}{\rho_{DE}}=-1+\frac{2\dot{H}(f'-1)-H\dot{f'}+\ddot{f'}}{-3H\dot{f'}+3(H^{2}+\dot{H})(f'-1)-(f-R)/2}\,.
\end{equation}

%%%%%%%%%%%%%%%%%%%%%%%%%%%%%%%%%%%%%%%%%%%%%%%%%%%%%%%%%%%%%%%%%%
\subsection{Numerical calculation}

We solve equation \eqref{eq:math:ex9} numerically in the $f(R)$ model
\eqref{eq:math:ex5}. To find the exact value of $R_{s}$, we require that
the density parameter for non-relativistic matter equals $\Omega_{m}=0.3$ and the
Hubble parameter is $H_{0}=72$ km/s/Mpc in the present
Universe.

\begin{figure}[!htb]
\centering
\includegraphics[keepaspectratio,width=9cm]{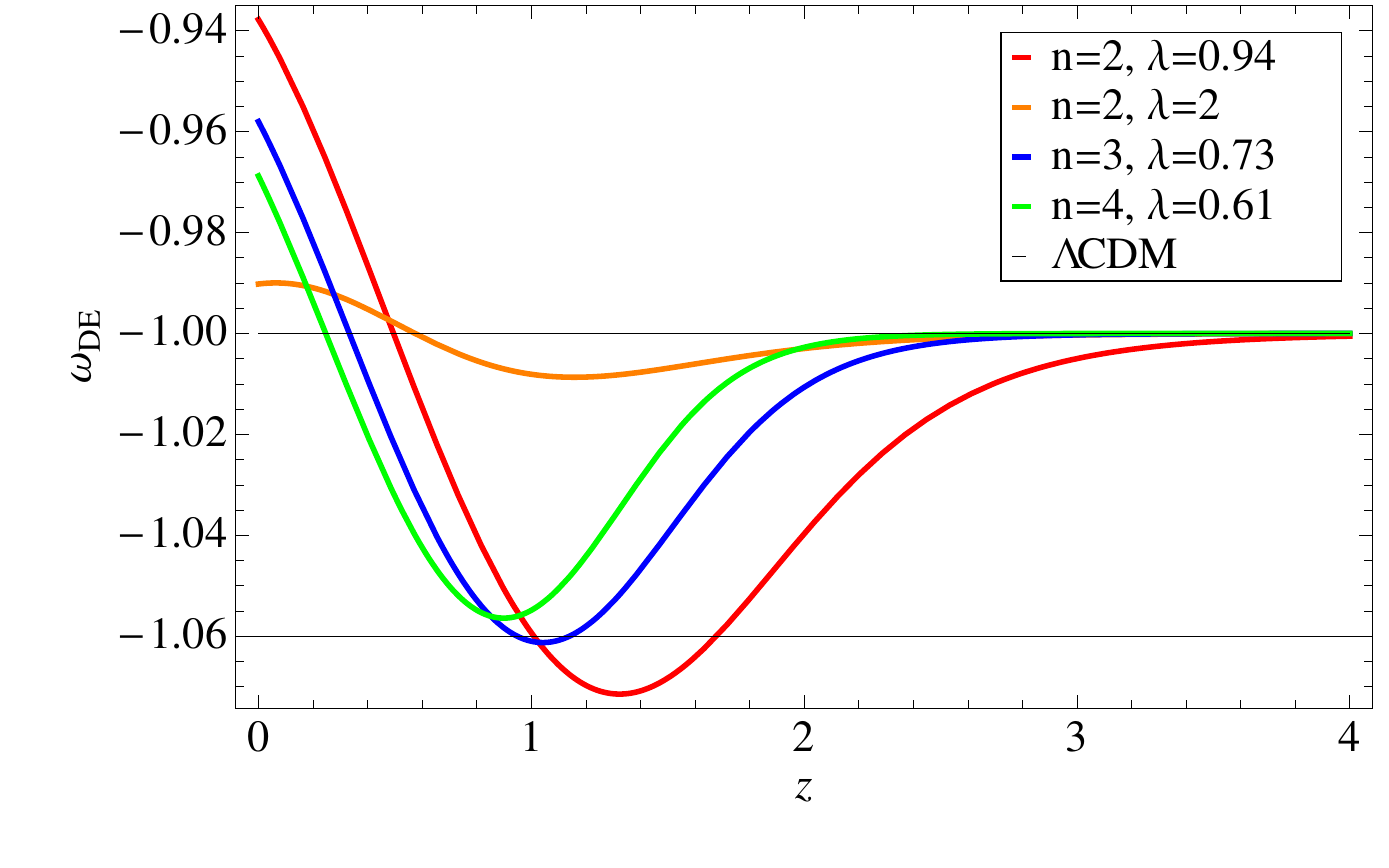}
\caption{\label{fig:1} Evolution of the equation-of-state
parameter $\omega_{DE}$ for DE in the $f(R)$ model \eqref{eq:math:ex5}.}
\end{figure}

Figure~\ref{fig:1}
depicts the evolution of $\omega_{DE}$ as a function of
redshift $z$ for reference values of $n$ and $\lambda$. Remarkably,
the condition of stability of the future de Sitter asymptotic solution~\cite{deSitterstab} imposes constraints on the free parameter $\lambda$. For instance, if $n=2$ then $\lambda>0.94$,  if $n=3$ then $\lambda>0.73$, and if $n=4$ then $\lambda>0.61$ according to~\cite{lambdaconstr}. The parameter $\omega_{DE}$ approaches the constant value $\omega_{DE}=-1$ as we increase $\lambda$ for fixed $n$, that is in
the $\Lambda$CDM-like limit. For minimal
allowed values of $\lambda$, deviation of the DE equation-of-state
parameter from the value $\omega_{DE}=-1$
for redshifts $z\lesssim 2$ is consistent with recent observational
data~\cite{exprerimentw}.

For the background metric, we find that the phantom boundary crossing
($\omega_{DE}=-1$) occurs at small redshifts $z\lesssim1$. This phantom
crossing is not peculiar to the specific choice of the function
\eqref{eq:math:ex5}. It is a generic feature of all models which
obey $f''(R)>0$.\par

%%%%%%%%%%%%%%%%%%%%%%%%%%%%%%%%%%%%%%%%%%%%%%%%%%%%%%%%%%%%%%%%
\subsection{The First Iteration Approach}
\label{subsec:fonfirst}

Numerical solution of eq.\,\eqref{eq:math:ex9} is not convenient for implementation in the computer simulation programme MGCAMB we would like to use. An alternative way is to use for the MGCAMB variables the expressions derived within the $f(R)$ gravity (i.e. eq.\,\eqref{eq:math:ex13} for $\omega_{DE}$ instead of $\omega_{DE}=-1$,
etc.) with the scale factor $a(t)$
solving the $\Lambda$CDM equations. It is the so-called iteration
method~\cite{Starmodel}. We consider the first iteration only and call
it the First Iteration Approach (FIA). It allows us to catch the
leading deviation of background evolution in the $f(R)$ model from
that in the $\Lambda$CDM model (that it is also necessary to determine the
change in the Integrated Sachs--Wolfe effect~\cite{ISW}) and compute
matter density perturbations in Sec.~\ref{sec:cosmo} more precisely.

\begin{figure}[!htb]
\centering % \begin{center}/\end{center} takes some additional vertical space
\includegraphics[keepaspectratio,width=9cm]{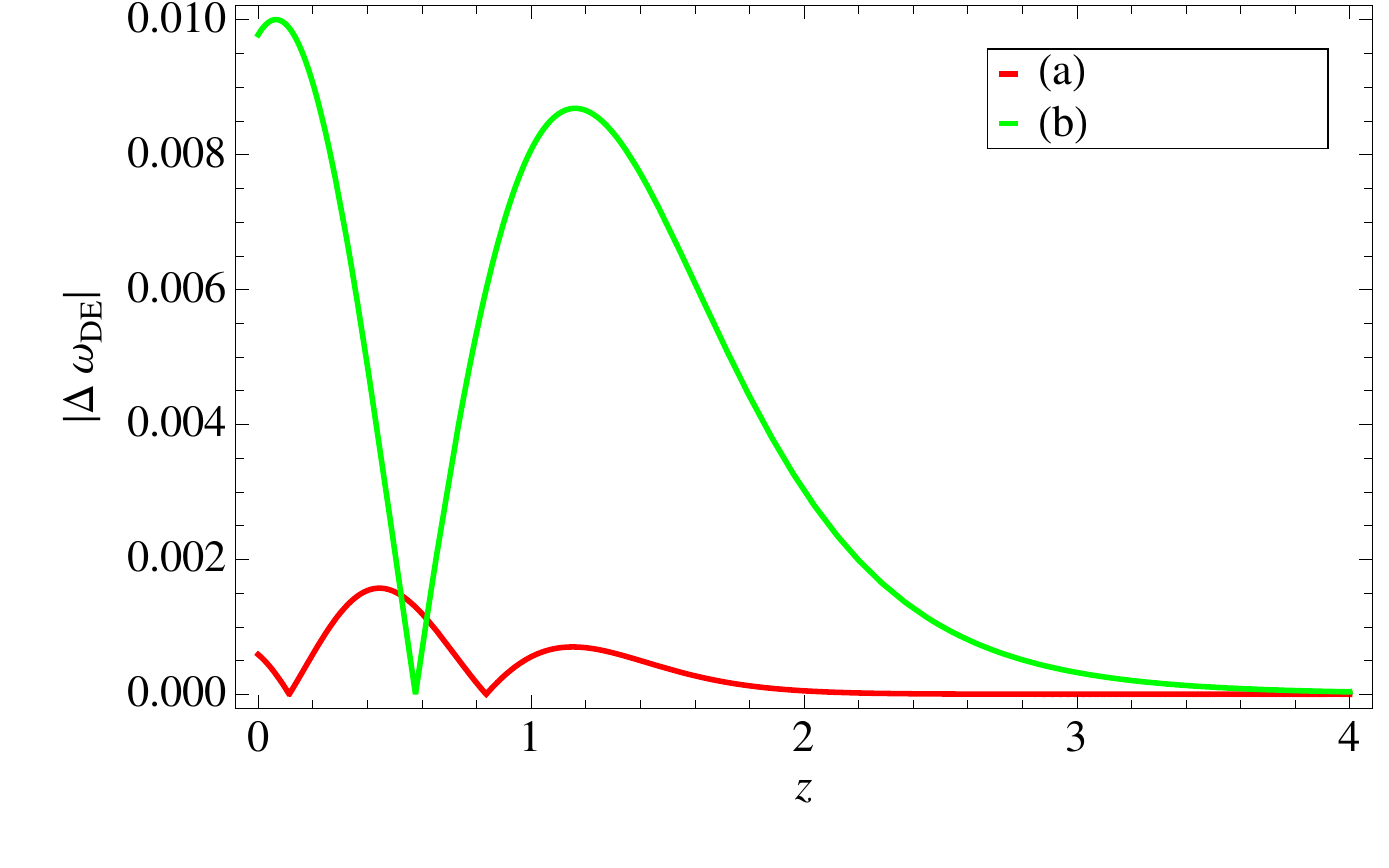}
\caption{\label{fig:2} Modulus of deviation of the
  equation-of-state parameter $\omega_{DE}$ derived by the FIA
from that derived by numerical calculation (a),
  modulus of deviation of the constant value $\omega_{DE}=-1$ inherited
  by  the $\Lambda$CDM model from the precise calculation
  of the equation-of-state parameter $\omega_{DE}$ (b).}
\end{figure}

Curve (a) in Fig.~\ref{fig:2}
represents the modulus of deviation of the equation-of-state parameter $\omega_{DE}$ for DE derived by
FIA from that derived by the straightforward numerical calculation
described above for $n=2$, $\lambda=2$ and the same values of
$\Omega_m$ and $H_0$ (the same present epoch) adopted in
Fig.~\ref{fig:1}. The deviation do not exceed
$0.2$\,\%. Obviously, it is even smaller for larger values of $n$ or
$\lambda$. Curve (b) in Fig.~\ref{fig:2} depicts the modulus of deviation
of the constant value $\omega_{DE}=-1$ in the $\Lambda$CDM model from the numerical calculation of the equation-of-state parameter
$\omega_{DE}$ for the same values of parameters and the same present
epoch. Clearly, at $\lambda=2$ the FIA yields a more accurate estimate of
$\omega_{DE}$ than one gets adopting the background behaviour of the $\Lambda$CDM model.  Moreover,
for values $\lambda>1.1$ such approach works better than the
approximation
within $\Lambda$CDM. Actually, for values $\lambda>1.5$
the deviation of the FIA results from the numerical solution is less than 1\,\%.
This precision is enough to extract values of cosmological
parameters with the percent accuracy, so we adopt it in what follows.

%%%%%%%%%%%%%%%%%%%%%%%%%%%%%%%%%%%%%%%%%%%%%%%%%%%%%%%%%%%%%%%%%%%%%%%%%%%%%
\section{Matter perturbations}
\label{sec:fluct}

%%%%%%%%%%%%%%%%%%%%%%%%%%%%%%%%%%%%%%%%%%%%%%%%%%%%%%%%%%%%%%%%%
\subsection{Linear evolution of the matter perturbations}

We turn to the linear evolution of matter density fluctuations in cosmological models based on $f(R)$ gravity. We define metric perturbations by
\begin{equation}  \label{eq:math:ex14}
ds^{2}=-(1+2\Phi)dt^{2}+a^{2}(t)(1-2\Psi)\delta_{ij}dx^{i}dx^{j}\,.
\end{equation}
For sub-Hubble modes in the quasi-static approximation, the equations for
the matter density contrast $\delta\equiv \delta \rho/\rho$ and the
gravitational slip in $f(R)$ gravity read~\cite{FT2010}
\begin{equation}  \label{eq:math:ex15}
\ddot{\delta}+2H\dot{\delta}-4\pi G_{eff}(t,k)\rho\delta=0\,,
\end{equation}
\begin{equation}  \label{eq:math:ex16}
\frac{\Psi}{\Phi}=\eta(t,k)\,,
\end{equation}
where
\begin{equation}  \label{eq:math:ex17}
G_{eff}(t,k)\equiv\frac{G}{f'}\frac{1+4\frac{k^{2}}{a^{2}}\frac{f''}{f'}}{1+3\frac{k^{2}}{a^{2}}\frac{f''}{f'}}\,,
\end{equation}
\begin{equation}  \label{eq:math:ex18}
\eta(t,k)\equiv\frac{1+2\frac{k^{2}}{a^{2}}\frac{f''}{f'}}{1+4\frac{k^{2}}{a^{2}}\frac{f''}{f'}}\,.
\end{equation}

In the quasi-GR regime ($f'\approx1$), the effective scalaron mass is given by~\cite{FT2010}
\begin{equation}  \label{eq:math:ex19}
M_{s}^{2}\approx \frac{1}{3f''(R)}\,.
\end{equation}
There are two different extreme regimes of evolution of the density
fluctuations: $M_{s}\gg k/a$ and $M_{s}\ll k/a$. The former regime
corresponds to $G_{eff}\approx G$ (thus, the evolution of $\delta$ mimics
that in GR), whereas the latter corresponds to $G_{eff}\approx 4G/3$
(this is the case of amplification of matter density perturbations
mentioned in Introduction).  Consequently, the effective gravitational
"constant" increases up to $33\,\%$, independently of the
functional form of $f(R)$.

%%%%%%%%%%%%%%%%%%%%%%%%%%%%%%%%%%%%%%%%%%%%%%%%%%%%%%%%%%%%%%%%%%%
\subsection{Beyond the linear regime and LGC}
\label{subsec:fluctLGC}

Matter perturbations of characteristic scale $\ell$ exhibit an extra
growth\,\cite{chameleonmech} for small
$\ell<\lambda_{c}\,(R=\kappa^{2}\rho_{\ell})$, where
$\lambda_{c}=M_{s}^{-1}$ is the Compton wavelength of the scalaron
field, $\rho_{\ell}$ is the local energy density of a particular
structure. At smaller scales referring to given astrophysical objects,
the local tests of $f(R)$ gravity become important.

Scalaron has to be heavy enough and unobservable to pass both
the cosmological and the Solar system tests. The chameleon mechanism takes care of
LGC for compact objects in the present Universe. Scalaron mass $M_{s}$
depends on a local value of the Ricci scalar $R$ arising in a local structure of the characteristic size $\ell$. Cavendish-type experiments with the
values $\rho_{\ell}\sim 10^{-12}$ g/cm$^{3}$ and $\ell\sim10^{-2}$
cm~\cite{Cavendishold} give a fairly weak constraint on model
parameters $n$, $\lambda$. In the case of Solar system tests, for
minimal allowed value $\rho_{\ell}\sim 10^{-24}$ g/cm$^{3}$ and
length-scale $\ell=1$ Au$=1.5\times 10^{13}$ cm we obtain the stronger
constraint $n\geq2$. It seems that investigating larger objects --
galaxies and galaxy clusters -- one can impose much stronger constraints
because the energy density on the outskirts of halos can be
estimated by the present cosmic background matter density
$\rho_{\ell}\sim10^{-29}$ g/cm$^{3}$. Really, the involved linear
effects begin to dominate on scales with such matter
density. Therefore we are interested in a violation of the Compton
condition $\ell<\lambda_{c}$ for linear perturbations, since galaxy tests do
not lead to tighter constrains. Moreover, as it was
shown in~\cite{chameleonmech}, the thin shell bound for a galaxy is
overly restrictive. To summarize different LGC, the range
$n\geqslant2$ is already sufficiently large to pass the
Solar system and other tests.

On the other hand, if we want to have an additional growth of linear
perturbations in the present Universe, we should put $M_{s}\gtrsim
H_{0}$ which corresponds to the Compton wavelength less but not much
less than the size of a visible part of the Universe. This condition
together with the aforementioned LGC give finally the constraint $n\gtrsim2$. Indeed,
in the case $n=2$ for the minimal available value $\lambda=0.94$, we obtain $M_{s}/H_{0}=3.6$ at the present epoch described by $\Omega_{m}=0.3$, $H_{0}=72$ km/s/Mpc.

%Generally, the range $n\gg 2$ is acceptable but with less pronounced effect of the extra growth on these scales.

It is convenient to use the dimensionless Compton wavelength squared
in Hubble units today $B_{0}=\left.(f''/f')(dR/d\ln
H)\right|_{t=t_{0}}$~\cite{B0}, the so-called deviation index, which
plays an important role in the both cosmological and local tests of $f(R)$
models. We calculate this index for different values of $\lambda$ at
fixed $n=2$ for the same present epoch described in the previous paragraph. We find $B_{0}=1.92\times10^{-1}$, $5.8\times10^{-5}$,
$2.4\times10^{-5}$ and $1.5\times10^{-6}$ for $\lambda=0.94$, $8$,
$10$ and $20$, respectively. Noteworthy, we calculate $B_{0}$ within
the gravity law~\eqref{eq:math:ex5} which corresponds to the scalaron mass
squared behavior $M_{s}^{2}\propto R^{6}$, whereas for the often used
Bertschinger--Zukin (BZ) parametrization one has
$M_{s}^{2}\propto R^{2}$~\cite{BZ} for large values of the scalar
curvature $R$, so that the effective cosmological constant grows logarithmically with $R$.\par

%For instance, there is not any visible density perturbation enhancement in linear regime for $n\geqslant13$.

%%%%%%%%%%%%%%%%%%%%%%%%%%%%%%%%%%%%%%%%%%%%%%%%%%%%%%%%%%%%%%%%%%%%%
\section{Parameter constraints}
\label{sec:cosmo}

We carry out the Markov Chain Monte Carlo (MCMC) analysis for
the $\Lambda$CDM model and the $f(R)$ gravity described by \eqref{eq:math:ex5}
with and without one massive sterile neutrino which is taken to be thermalized and shares the
same temperature as the active neutrinos. We neglect masses of
the three standard neutrinos  compared to that of one sterile
neutrino. We have modified the MGCAMB~\cite{MGCAMB1,MGCAMB2} that
allows to implement $f(R)$ gravity by adopting \eqref{eq:math:ex17}
and \eqref{eq:math:ex18}. We change the background evolution equations
as provided by the FIA (see
subsection~\ref{subsec:fonfirst}) catching the first nonvanishing
correction to the background evolution in the $\Lambda$CDM model. We plugged
the above modified MGCAMB code into CosmoMC~\cite{COSMOMC1,COSMOMC2}
to constrain the model parameters. We use six
standard free fitting parameters: the density parameters for baryon matter
$\Omega_{b}h^{2}$ and for cold dark matter without neutrino
$\Omega_{c}h^{2}$, the sound horizon angle
$\theta_{*}\equiv100r_{s}/D_{A}(z_{*})$, the optical depth $\tau$, the
scalar spectral index $n_{s}$ and the amplitude of the primordial
power spectrum $\Delta_{\mathcal{R}}^2$. Occasionally, we add extra
fitting parameters such as the total mass of three active neutrinos
$\sum m_{\nu}$ or the mass of one sterile neutrino
$m_{\nu,\,\rm{sterile}}$, so that the relative contribution of the dark
matter to the present energy density is $\Omega_{DM}=\Omega_{c}+\Omega_{\nu}$.

When we work with modified gravity, we consider $\lambda$ as a fitting
parameter. We fix another parameter of $f(R)$
gravity as $n=2$ because it is the minimal integer value for which the
effect of the density perturbation enhancement in the linear regime is
the most pronounced (see subsection~\ref{subsec:fluctLGC}).

Performing numerical calculations, we find the regions of parameter space
consistent with cosmological data at the 65\,\% and 95\,\% confidence levels
and outline them on plots presented below.

%%%%%%%%%%%%%%%%%%%%%%%%%%%%%%%%%%%%%%%%%%%%%%%%%%%%%%%%%%%%%%%
\subsection{Cosmological Data}
\label{subsec:cosmodata}

In our analysis we use different data sets. The first set is the
measurement of the CMB temperature power spectrum from the one-year data
release of the Planck satellite~\cite{PLANCK} supplemented with the
low-$\ell$ polarization measurements from the nine-years observations
of the WMAP satellite~\cite{PLANCKlowl1,PLANCKlowl2}. This data set is
designated below as 'Planck'. We extend this data set with CMB
measurements at high-$\ell$ by the Atacama Cosmology Telescope
(ACT)~\cite{ACT} and the South Pole Telescope
(SPT)~\cite{SPT1,SPT2,SPT3}. We refer to these measurements below as $ePlanck$.

We also include different measurements of baryon acoustic oscillation
(BAO) including the LOWZ~\cite{LOWZ} and CMASS~\cite{CMASS} samples of
BOSS corresponding to SDSS DR11
%(expected in December 2014)
in the redshift range $0.15<z<0.43$ and $0.43<z<0.7$, respectively,
and also the 6dF Galaxy Survey~\cite{6dFGS} corresponding to
$z=0.106$. These BAO data sets do not overlap, and therefore
we can use them together. We refer to this set combination as $BAO$.

%Additionally, we add new data of baryon acoustic oscillation in the flux-correlation function of the Ly-$\alpha$ forest of high-redshift quasars. Because of the small %statistical significance, the most BAO analyses in the past reported a measurement of the single value $D_{V}/r_{d}$ where $D_{V}$ is the so-called %volume-averaged effective distance $D_{V}=[(1+z)D_{A}]^{2/3}[zD_{H}]^{1/3}$, $D_{A}$ is the angular distance, $D_{H}=c/H$, $H$ is expansion rate,
% $r_{d}$ is the sound horizon at the drag epoch (baryon decoupling from photons). This is not the optimal way to report the measurement, specially at high redshift. %In~\cite{BAOlya} parameter combinations $D_{A}/r_{d}$ and $D_{H}/r_{d}$ were measured simultaneously from the two-dimensional 2-point correlation %function at average absorber redshift $\langle z\rangle=2.34$. Knowing these values and its uncertainties with correlation coefficient between the two measurements %$r_{eff}=-0.43$ reported by Andreu Font-Ribera (afont@lbl.gov) we find $D_{V}/r_{d}=31.22\pm1.10(1\sigma)$, which is used in COSMOMC package. %However, combining two measurements to a single value we lose some information.

In addition, we use the Hubble constant measurement~\cite{HST}
mentioned below as $H_{0}$ and the full-sky lensing potential
map~\cite{LENSING} called $LENS$.\par

Finally, we use observations of galaxy clusters. Data on
cluster mass function measurements are taken
from~\cite{CLUSTERS1,CLUSTERS2} using the likelihood data described
in~\cite{CLUSTERS3}. In this study, a sample of 86 massive galaxy
clusters in the ranges
$z<0.2$ and $z\approx0.4-0.9$ with masses measured with
about 10\,\% accuracy by the  \emph{Chandra} X-ray telescope was used to
determine the cluster mass function (the subsample of distant massive clusters was
taken from the \emph{400d} X-ray galaxy cluster survey~\cite{CLUSTERS4}). Likelihoods were obtained for the DE model with a constant in time parameter $\omega_{DE}$. We take the results for
$\omega_{DE}=-1$. Deviation of an effective time-dependent $\omega_{DE}$
from the value $\omega_{DE}=-1$ is below 0.1\,\% for $\lambda>3.6$. As is shown below, this choice of $\lambda$ is justified. We refer to this data as $CL$. \par

%The tightest constraints on cosmological parameters are obtained when all above data are included in our analysis. We refer to this case as 'Full' combination.

\subsection{Results and discussion}
\label{subsec:result}

\begin{figure}[tbp]
\centering % \begin{center}/\end{center} takes some additional vertical space
\includegraphics[keepaspectratio,width=3.7cm]{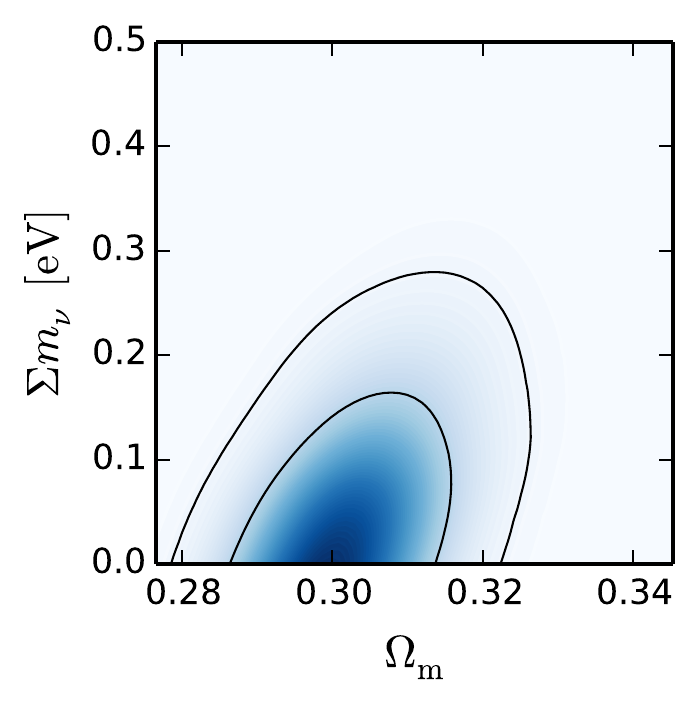}
\includegraphics[keepaspectratio,width=3.7cm]{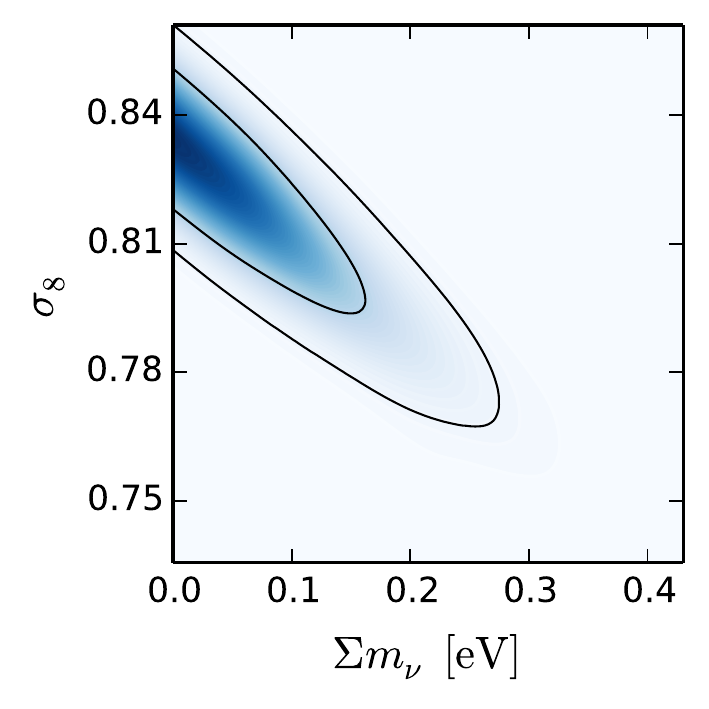}
\includegraphics[keepaspectratio,width=3.7cm]{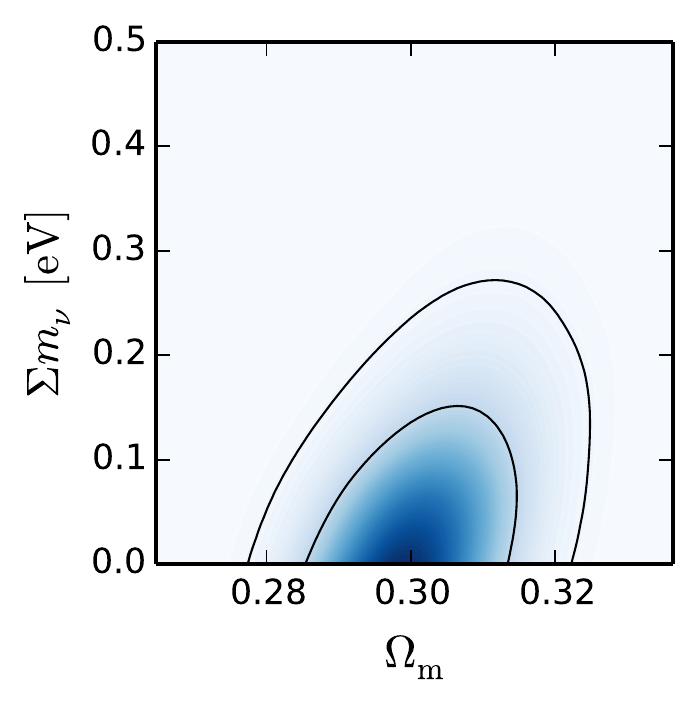}
\includegraphics[keepaspectratio,width=3.7cm]{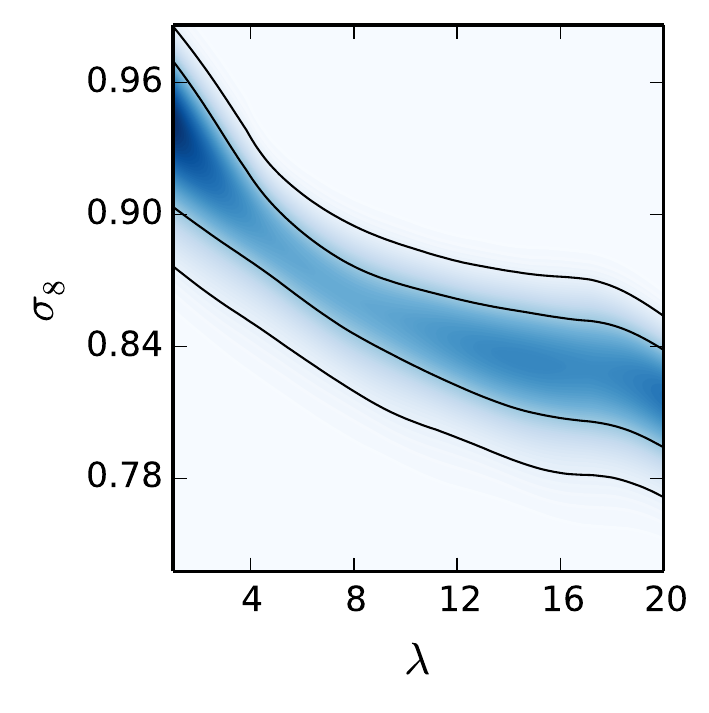}
\caption{\label{fig:3} Constraints for the $\Lambda$CDM model in the $\sum
  m_{\nu}$-$\Omega_{m}$, $\sigma_{8}$-$\sum m_{\nu}$ planes (two left panels)
 and for $f(R)$ gravity in the $\sum m_{\nu}$-$\Omega_{m}$,
 $\sigma_{8}$-$\lambda$ planes (two right panels) assuming one
massive and two massless active neutrinos
within the $ePlanck$+$BAO$+$LENS$+$H_{0}$ data set.}
\end{figure}

At first, we compare the $\Lambda$CDM and $f(R)$ models with three active
neutrinos (among them only one is taken massive) using the
$ePlanck$+$BAO$+$LENS$+$H_{0}$ data set. From Fig.~\ref{fig:3} we see that
$f(R)$ gravity does not relax the upper limit on neutrino mass as compared to
the $\Lambda$CDM model. On the rightmost panel in Fig.~\ref{fig:3} we see the
effect of $f(R) $ gravity on the growth of matter density
perturbations. Clearly, the enhancement of the $\sigma_{8}$ value is
not restricted because the data set that constrains the structure formation is
not included yet. A notable feature of modified gravity consists in
an apparent peak of the posterior probability distribution at low values of
$\lambda$ which corresponds to the most probable growth rate of the
structure formation. The reason is that the sensitivity of the CMB multipole spectrum to $f(R)$ models is mainly due to the late ISW effect changing low multipoles according to \cite{Planck_ISW1,Planck_ISW2}. Then the strong modification of
the density perturbation evolution on linear scales provides better parameter
convergence. This tendency is in full compliance with the recent results
of \cite{B0last}. Moreover, we reproduce the neutrino mass constraint from
that article in the limit of negligible deviation from GR (high values of $B_{0}$) according to the first and the third panels in Fig.~\ref{fig:3}. Remarkably, here the strong constraint on neutrino mass is obtained irrespective of the evolution history of linear perturbations.\par

%Galaxy cluster mass functions  constrains parameter combination $\sigma_{8}(\Omega_{m}/0.25)^{0.47}$

\par

%Because of high probability of region with low value of $\lambda$ (point with $\lambda<1.5$ are located in $1\sigma$ confident region), we keep background evolution in the $\Lambda$CDM model.

\begin{figure}[tbp]
\centering % \begin{center}/\end{center} takes some additional vertical space
\includegraphics[keepaspectratio,width=3.7cm]{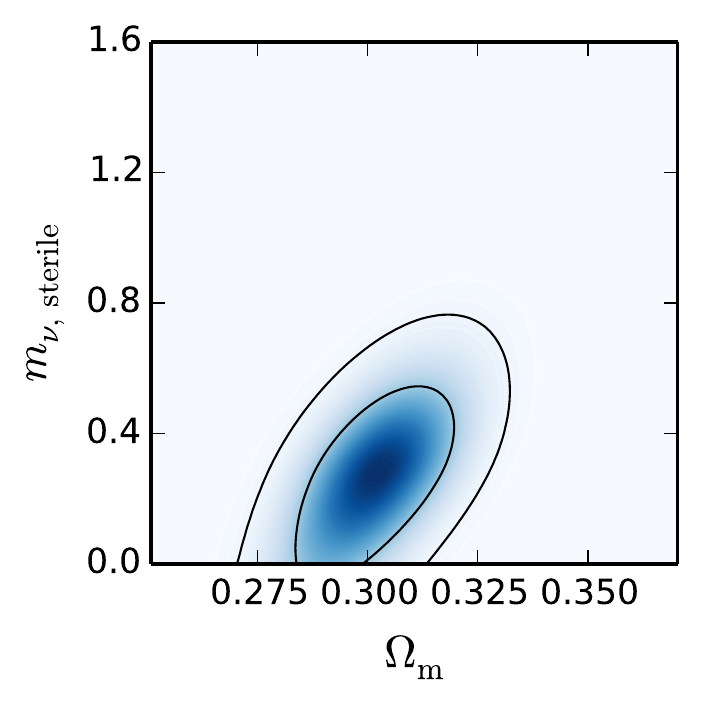}
\includegraphics[keepaspectratio,width=3.7cm]{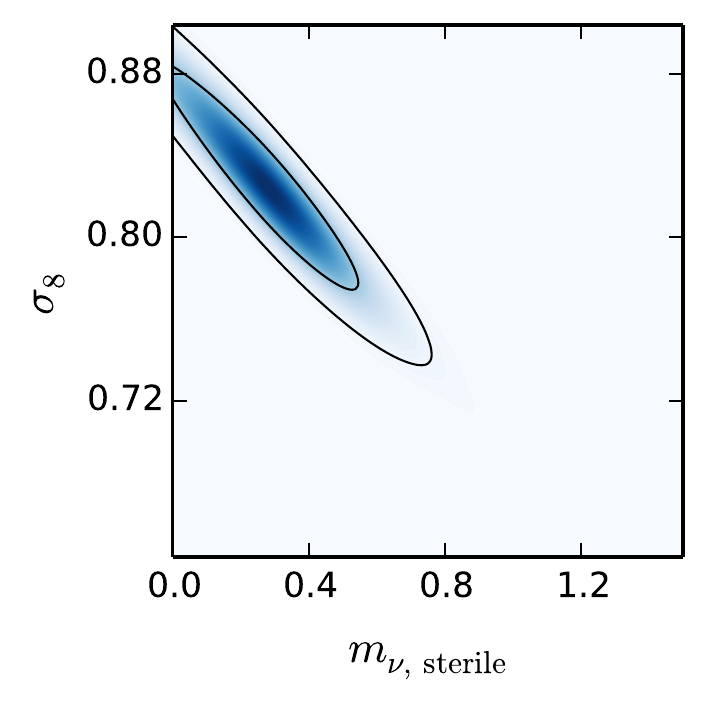}
\includegraphics[keepaspectratio,width=3.7cm]{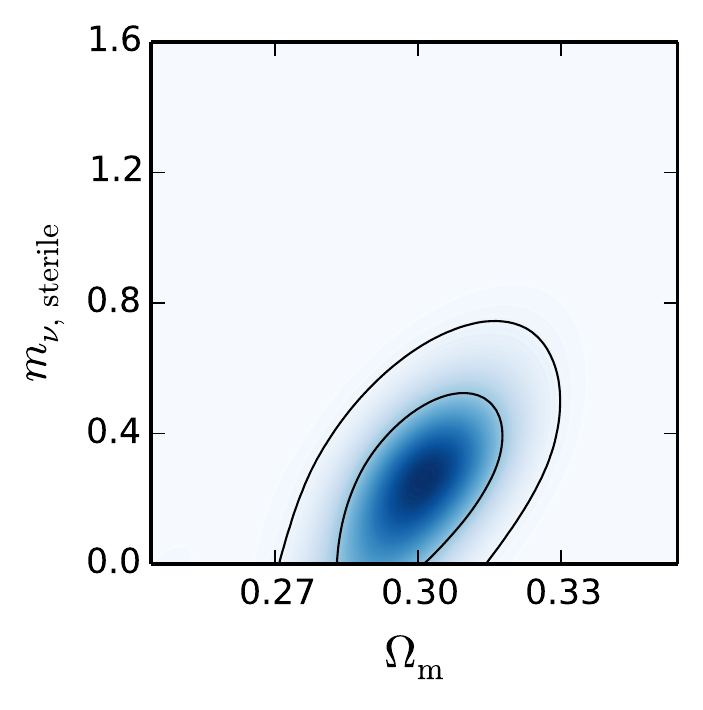}
\includegraphics[keepaspectratio,width=3.7cm]{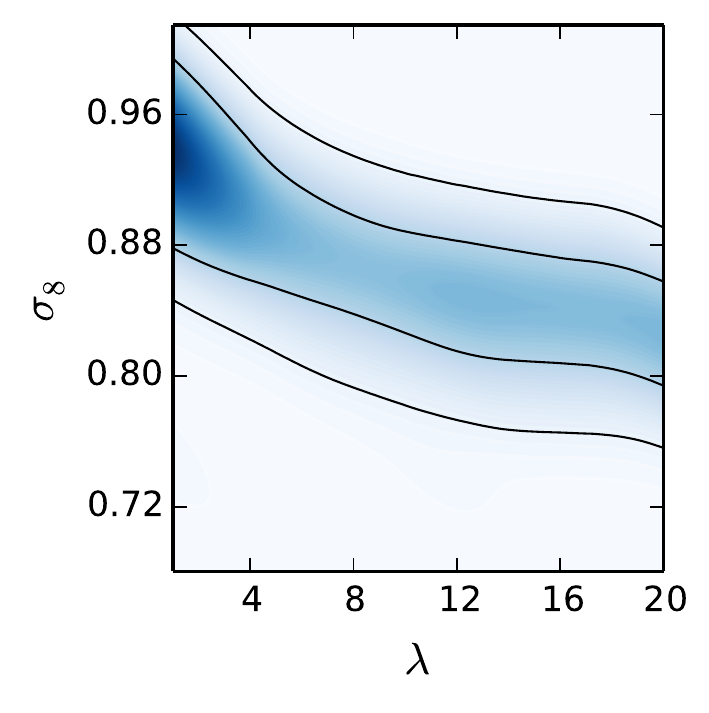}
\caption{\label{fig:4}
Constraints for the $\Lambda$CDM model in the
$m_{\nu,\,\rm{sterile}}$-$\Omega_{m}$,
$\sigma_{8}$-$m_{\nu,\,\rm{sterile}}$ planes (two left panels) and
for $f(R)$ gravity in the $m_{\nu,\,\rm{sterile}}$-$\Omega_{m}$,
$\sigma_{8}$-$\lambda$ planes (two right panels) assuming one massive
sterile and three massless active neutrinos within the $ePlanck$+$BAO$+$LENS$+$H_{0}$ data set.}
\end{figure}

In principle, the introduction of massive sterile species can improve the situation
and leads to a weaker mass constraint. The results are shown in
Fig.~\ref{fig:4} for different models: $\Lambda$CDM and $f(R)$
gravity. Indeed, the value of (sterile) neutrino mass increases as
compared to the case with only three active neutrinos. This can be
understood as a necessity to keep a constant redshift of the matter-radiation equality, $z_{eq}$ which is determined from CMB
observations as explained in~\cite{CAMB_zeq}. Nevertheless, the extra
growth of linear perturbations in modified gravity at low values of $R$ cannot be canceled by massive neutrinos, contrary to what was suggested in
\cite{Star_1eV}. The reason is that the sterile neutrino mass is
constrained quite strongly for any evolution history of the linear
perturbations by the CMB+BAO data only.

%Most likely, $f(R)$ gravity will not change this situation appreciably
%when we include information that constrains structure formation (or
%$\sigma_{8}$ equivalently). We ascertain this directly below.

% This can be explained by necessity to keep a constant redshift of
% matter-radiation equality, $z_{eq}$, which is measured from CMB
% observation as explained in~\cite{CAMB_zeq}

\par

\begin{figure}[tbp]
\centering % \begin{center}/\end{center} takes some additional vertical space
\includegraphics[keepaspectratio,width=3.7cm]{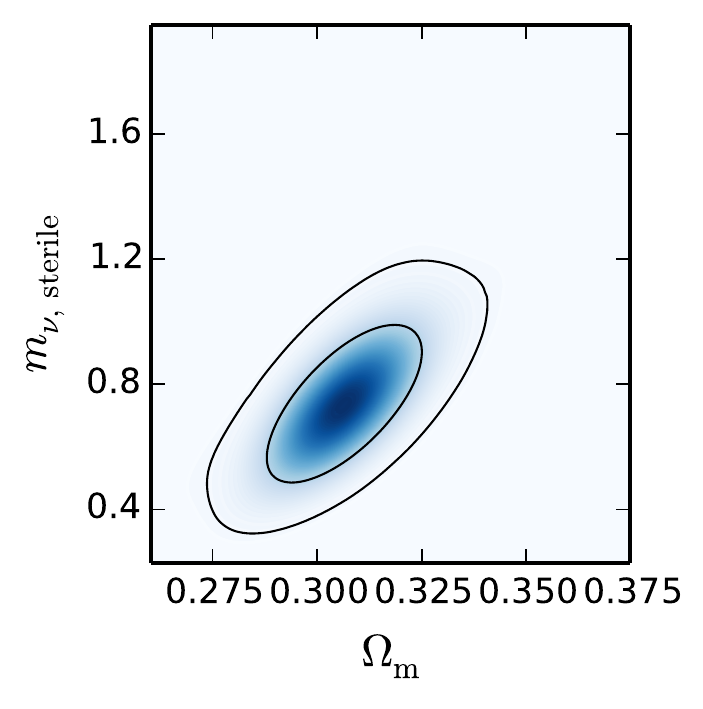}
\includegraphics[keepaspectratio,width=3.7cm]{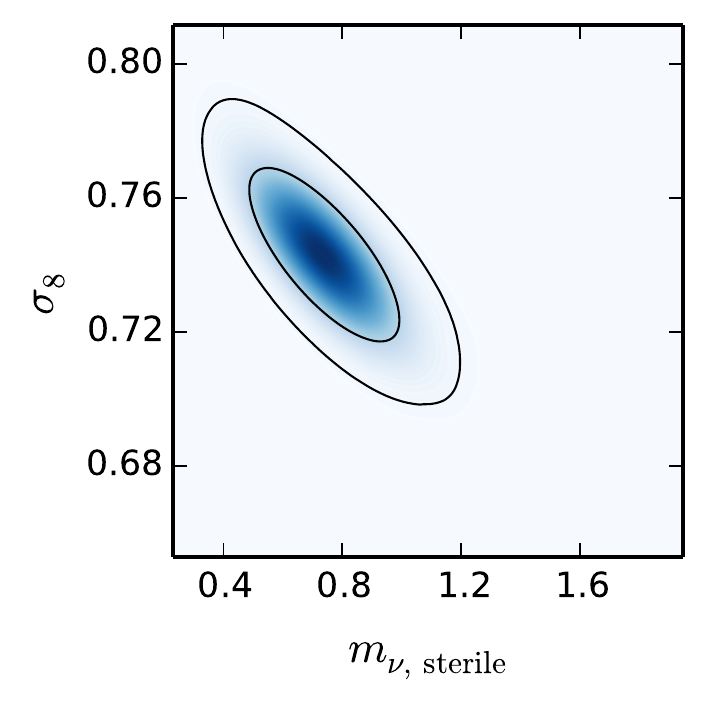}
\includegraphics[keepaspectratio,width=3.7cm]{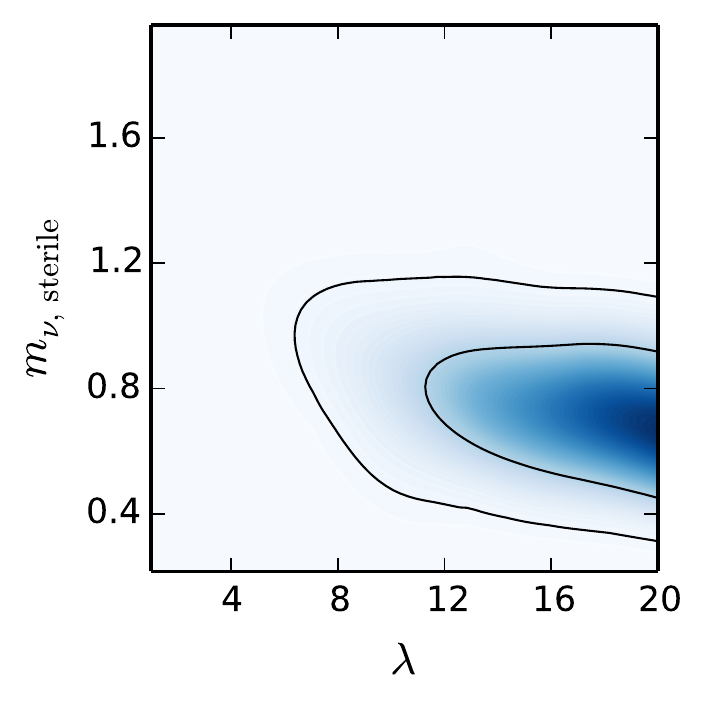}
\includegraphics[keepaspectratio,width=3.7cm]{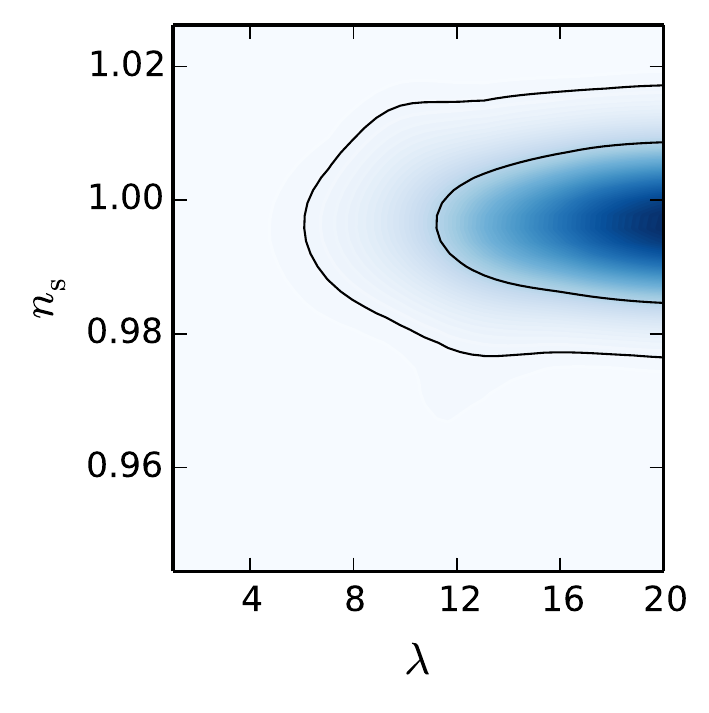}
\caption{\label{fig:5} Viable regions in the parameter space
  for $f(R)$ gravity (with background evolution according to the FIA in the
  $m_{\nu,\,\rm{sterile}}$-$\Omega_{m}$,
  $\sigma_{8}$-$m_{\nu,\rm{sterile}}$,
  $m_{\nu,\,\rm{sterile}}$-$\lambda$, $n_{s}$-$\lambda$
  planes (panels from left to right) assuming one massive sterile and three
  massless active neutrinos within the $ePlanck$+$BAO$+$LENS$+$H_{0}$+$CL$ data set.}
\end{figure}

In order to check the growth of $\sigma_{8}$,
%and to consider the
%growth rate of linear density perturbations in $f(R)$ model properly
we use the galaxy cluster data set. From Fig.~\ref{fig:5} we see that
the galaxy cluster data constrain $\sigma_{8}$ at lower values as compared
to the Planck CMB data \cite{Planck_XVI}.
% The matter is that galaxy cluster mass functions constrains
% parameter combination $\sigma_{8}\cdot(\Omega_{m}/0.25)^{0.47}$ and
% for higher value of $\Omega_{m}$ (see~\cite{Planck_XVI}) it requires
% lower value of $\sigma_{8}$.
This tension can be resolved with massive neutrinos introduced into
the cosmological model, which suppress the matter density fluctuations
growth through the free-streaming effect discussed above. The tension
was first observed with pre-Planck data \cite{BR13,Hou14}, and it is
even more prominent when the Planck CMB and SZ clusters data are used,
see, e.g., \cite{Planck_XVI,Planck_XX,Hamann13,Beutler14}. This
tendency is also reflected on the plot of the second panel in
Fig.~\ref{fig:5}.

\par

%Hereinafter, we use the First Iteration Approach (see Subsec.~\ref{subsec:fonfirst}) to describe background evolution in $f(R)$ gravity more precisely.

\begin{figure}[tbp]
\centering % \begin{center}/\end{center} takes some additional vertical space
\includegraphics[keepaspectratio,width=3.7cm]{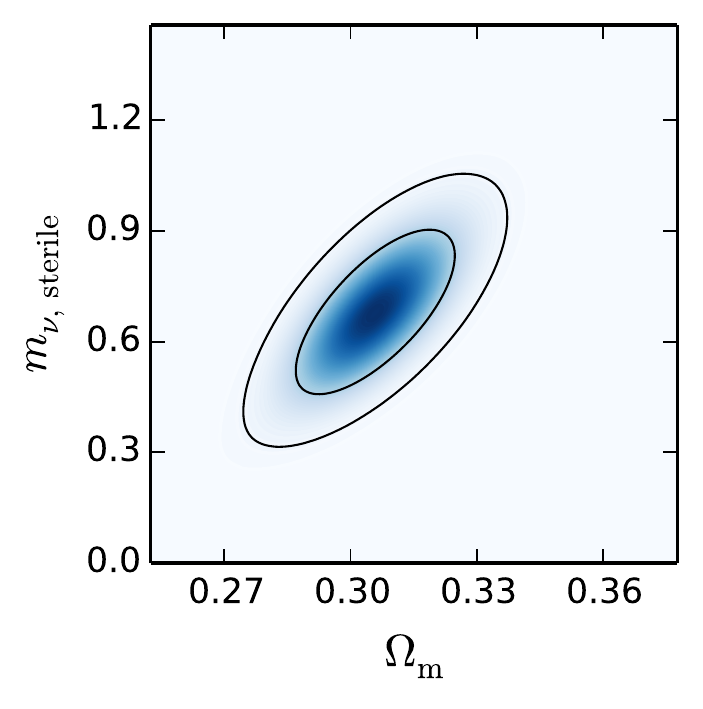}
\includegraphics[keepaspectratio,width=3.7cm]{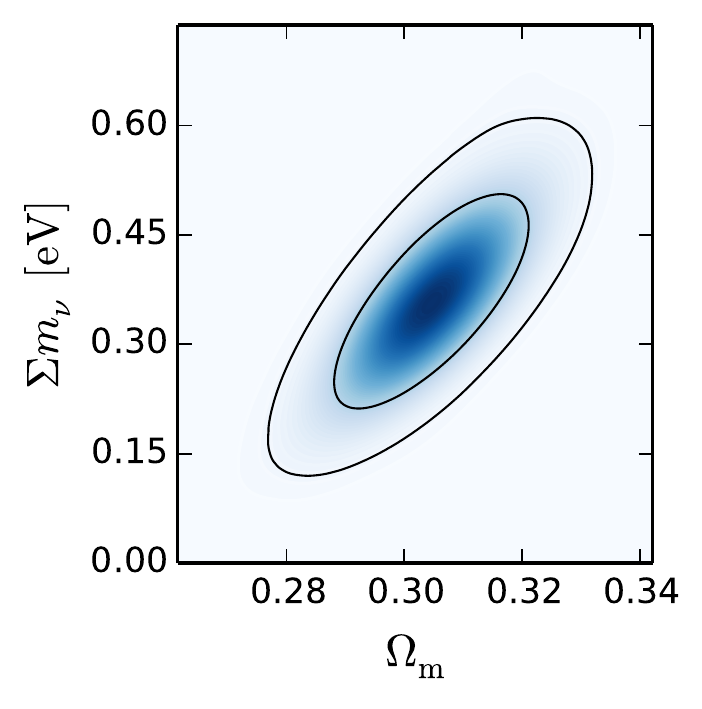}
\includegraphics[keepaspectratio,width=3.7cm]{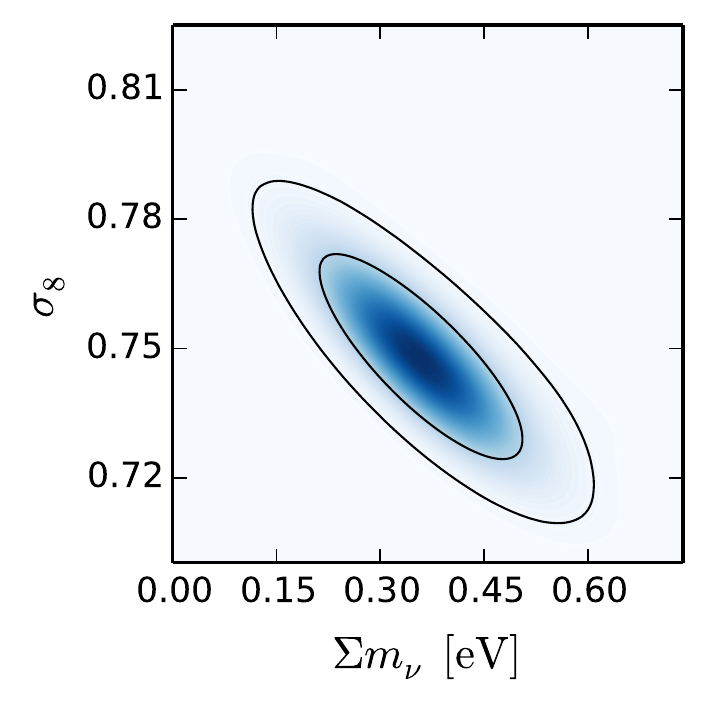}
\includegraphics[keepaspectratio,width=3.7cm]{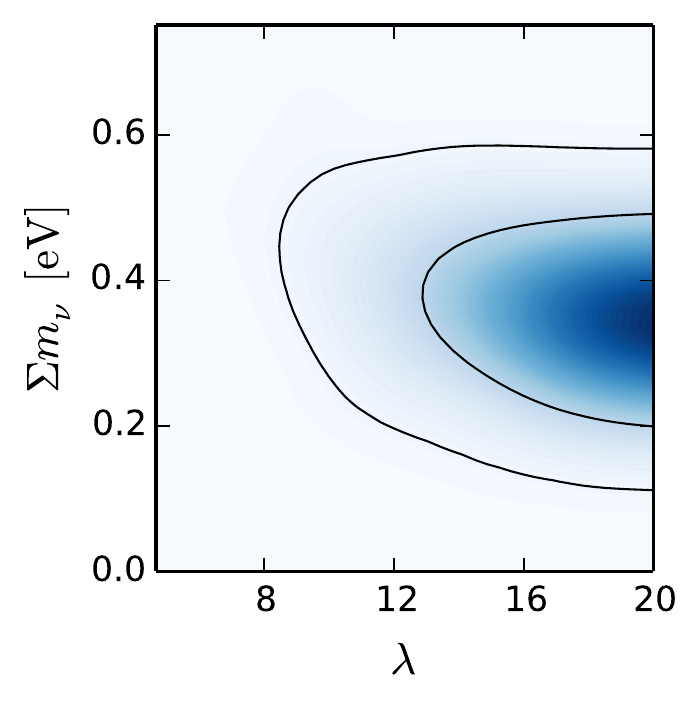}
\caption{\label{fig:66} Posterior distributions for
  $m_{\nu,\,\rm{sterile}}$ and $\Omega_{m}$ (the first panel)
  in the $\Lambda$CDM model with one massive sterile neutrino (assuming the active neutrinos are massless) and constraints on pairs $\sum
m_{\nu}$-$\Omega_{m}$, $\sigma_{8}$-$\sum
m_{\nu}$, $\sum m_{\nu}$-$\lambda$ (the other panels) in $f(R)$ gravity with one massive and two massless active neutrinos within the $ePlanck$+$BAO$+$LENS$+$H_{0}$+$CL$ data set.}
\end{figure}

According to the plot on the third panel in Fig.~\ref{fig:5} $f(R)$
gravity leads to slight degeneracy between sterile neutrino mass
and the single free parameter of the modified gravity model $\lambda$. While structures in the Universe grow faster for smaller $\lambda$, the value of sterile neutrino mass has to be increased to compensate for the extra growth of perturbations
at small scales in modified gravity. But the sterile neutrino mass is tightly
constrained from the $ePlanck$+$BAO$ data set. As a result, not much place
for the extra growth remains after the implementation of $f(R)$.  Indeed, the
marginalized constraint on the sterile neutrino mass within $f(R)$ gravity
is $0.47\,\,\rm{eV}$$\,<\,$$m_{\nu,\,\rm{sterile}}$$\,<\,$$1\,\,\rm{eV}$
(2$\sigma$) in contrast to
$0.45\,\,\rm{eV}$$\,<\,$$m_{\nu,\,\rm{sterile}}$$\,<\,$$0.92\,\,\rm{eV}$
(2$\sigma$) in $\Lambda$CDM with three active neutrinos taken
massless, according to the leftmost panels in Fig.~\ref{fig:5} and
Fig.~\ref{fig:66}. Our constraint on the sterile neutrino mass is
conservative because in reality the minimal sum of active neutrino
masses is nonzero -- it is either 0.05\,eV or 0.1\,eV corresponding to the normal
and inverted hierarchies.

\par

In addition, the rightmost panel in Fig.~\ref{fig:5} shows that after introducing one sterile neutrino $f(R)$ gravity does not change the
value of the scalar spectral index $n_{s}$ as compared to $\Lambda$CDM
\cite{Planck_XVI}. We note that the region of large values of
$\lambda$ corresponds approximately to the same structure formation as in
$\Lambda$CDM: the results presented in Fig.\,\ref{fig:5} match in this
limit similar results in $\Lambda$CDM.

Thanks to strong restrictions on $\sigma_{8}$ from galaxy cluster
observations, we can explore the function $f(R)$ by getting a constraint on the
model fitting parameter $\lambda$. Namely, from Fig.~\ref{fig:5} we find
$\lambda>9.4$ (2$\sigma$) in the case of the fourth massive sterile
neutrino and others taken massless. It implies a restriction on the
deviation index at present, $B_{0}<3.1\times10^{-5}$ (2$\sigma$) for
$\Omega_{m}=0.3$ and $H_{0}=72$ km/s/Mpc. When the
systematic uncertainty of the cluster mass function $\delta
M/M\approx0.09$ \cite{CLUSTERS2} is included in the likelihood
functions, the constraints are relaxed: $\lambda>8.2$ (2$\sigma$) and
$B_{0}<5.3\times10^{-5}$ (2$\sigma$). We can get restrictions on
$\lambda$ and $B_{0}$ in the Universe with the three active
neutrinos, too,  assuming one is massive: $\lambda>10.8$ (2$\sigma$) from
Fig.~\ref{fig:66} which corresponds to $B_{0}<1.8\times10^{-5}$
(2$\sigma$) without taking the systematic uncertainty of the galaxy
cluster measurements into account and $\lambda>9.6$ (2$\sigma$),
$B_{0}<2.8\times10^{-5}$ (2$\sigma$) with that. Using these constraints on the parameter $\lambda$, we check that
the use of FIA (see subsection~\ref{subsec:fonfirst}) and galaxy
clusters data set (see subsection~\ref{subsec:cosmodata}) are
justified. The obtained above bounds can be easily transformed to constraints on the present scalaron mass which determines the strength of $f(R)$ gravity.
For one massive sterile neutrino we obtain $M_{s}/H_{0}>194$
(2$\sigma$) with the cluster systematics and $M_{s}/H_{0}>255$ (2$\sigma$)
without it; in the case of three active neutrinos we get
$M_{s}/H_{0}>266$ (2$\sigma$) and $M_{s}/H_{0}>336$ (2$\sigma$),
respectively.

%that galaxy clusters are sensitive to $f(R)$ perturbation enhancement as well as neutrino erasement

We use the galaxy cluster data to get rid of degeneracy between
the massive sterile neutrino and modified gravity. The cluster mass function has
been extracted from the cluster data assuming the $\Lambda$CDM-like setup
for structure formation \cite{Tinker:2008ff}, so that the change in
gravity strength inherent in $f(R)$ model \eqref{eq:math:ex17}
remains unaccounted. In particular, the modified gravity impact on the
dynamics of matter streaming from infall region to the cluster halo in
recent epoch is neglected.  The median mass at all redshifts in the
galaxy clusters data set is near
$M_{500}=2.5\times10^{14}\,h^{-1}\,M_{\odot}$ that corresponds to
structures which were generated from the comoving critical scale
\cite{CLUSTERS1} $8\,h^{-1}$ Mpc. Such scale is rather big and $f(R)$
linear modification of density growth is quite important on these
scales. For example, cosmological macrostructures, such as filaments
which feed clusters with extra matter at $z\lesssim1$, have
moderate density contrast $\delta\rho/\rho\approx2-3$ and the effect of
modified gravity on this structure formation is quite noticeable (the
present size of cosmological filaments can be estimated as 10 Mpc). Therefore we have to find the range of $\lambda$ where the
approach used in \cite{Tinker:2008ff} is still valid.

\begin{figure}[tbp]
\centering % \begin{center}/\end{center} takes some additional vertical space
\includegraphics[keepaspectratio,width=9cm]{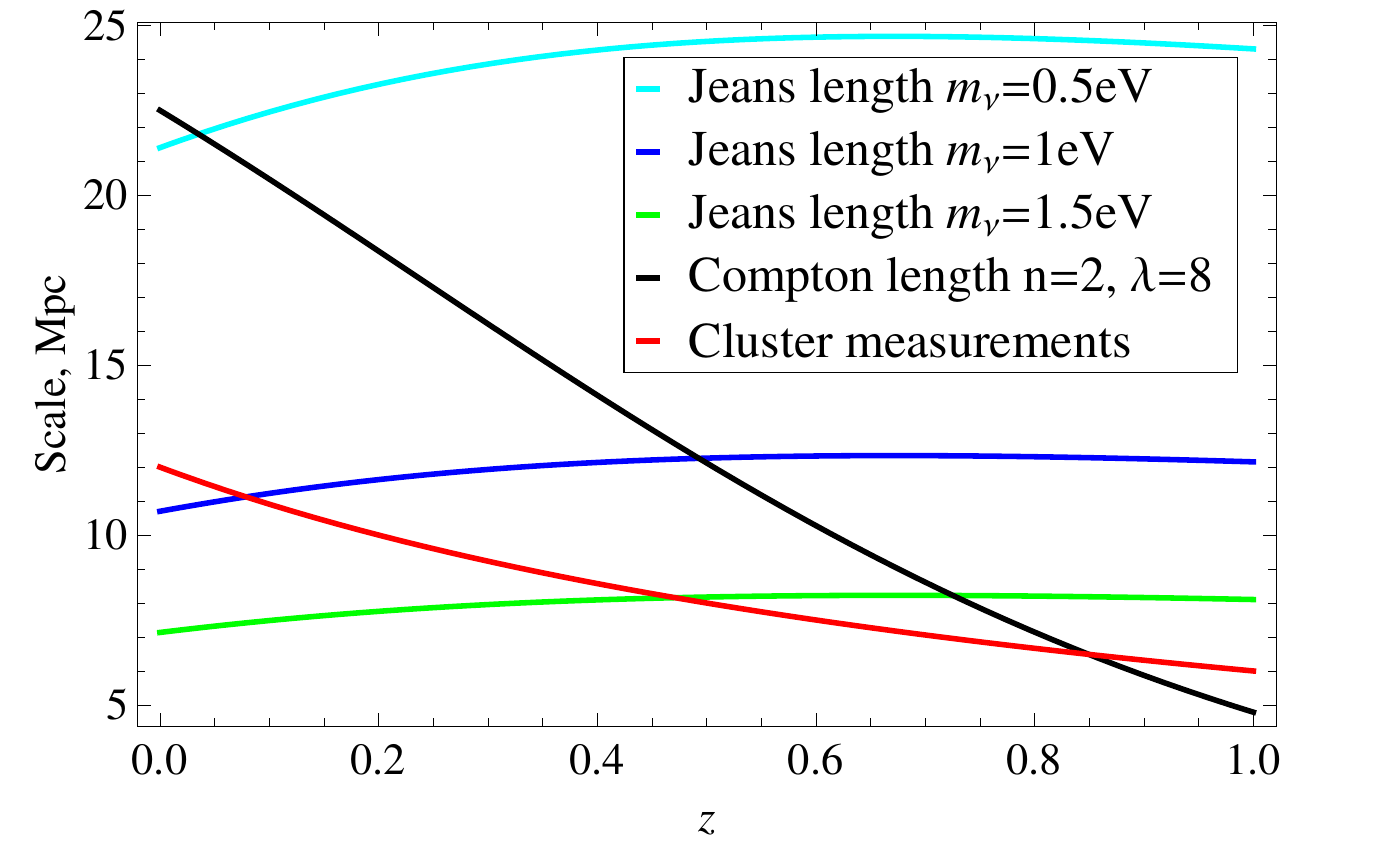}
\caption{\label{fig:6} The Jeans scale of neutrino free-streaming and
  the Compton length of scalaron ($n=2$, $\lambda=8$) as functions of
  redshift $z$ for various values of neutrino mass. Below the lines
  called {\it Jeans length}, perturbations are suppressed
  by the neutrino free streaming. Below the line named {\it Compton
    length}, the growth of perturbations is enhanced due to modified
  gravity. Clusters collect matter from regions of the size below
  the line labeled {\it Cluster measurements}.}
\end{figure}

According to $f(R)$ gravity, the growth of linear density perturbations is
extensive on scales below the Compton wavelength of the scalaron field with
the mass given by \eqref{eq:math:ex19}. On the contrary, light neutrinos
play restrictive role to $f(R)$ gravity by damping the  structure formation
most efficiently on
scales below the free-streaming (Jeans) length
\cite{freestreaming}
\begin{equation}  \label{eq:math:ex21}
\lambda_{FS}=7.7\frac{1+z}{\sqrt{\Omega_{\Lambda}+\Omega_{m}(1+z)^{3}}}\Big(\frac{1\,\rm{eV}}{m_{\nu}}\Big)h^{-1} \,\rm{Mpc}\,.
\end{equation}
In Fig.\,\ref{fig:6} 
we compare the Compton wavelength of
scalaron for the rather high value of $\lambda=8$, the free-streaming wavelength of
neutrino for various neutrino masses and the median critical size of
cluster which is used in the galaxy cluster measurements at low redshifts
$z<1$ assuming $\Omega_{m}=0.3$ and $h=0.72$. As clearly seen from
Fig.~\ref{fig:6}, for the galaxy cluster formation the
free-streaming effect is maximal during all evolution for all
interesting values of neutrino
masses. The situation with the $f(R)$ critical scale is more complicated. We see
that in the epoch $z\approx1$ when active formation of the largest cosmological
structures (galaxy clusters and filaments) begins, the
scalaron Compton length exceeds the
critical cluster size. Approximately from this time the $\Lambda$CDM approximation of galaxy cluster formation adopted for
galaxy cluster data in our case must be corrected for the increase of
gravity strength \eqref{eq:math:ex17}. This picture corresponds to $\lambda=8$. If we take modified gravity with more
pronounced effects, i.e. decrease $\lambda$, the Compton wavelength increases
too and the impact of gravity modification on the structure formation
starts earlier and extends to smaller structures.
If we describe the formation of galaxy
clusters with computer simulation properly accounting for the $f(R)$
gravity impact, the value of $\sigma_{8}$ has to be reduced because the number of
galaxy clusters in the Universe is fixed by the galaxy cluster data
set. For this reason, the constraint on parameter $\lambda$ obtained above
from the galaxy cluster mass function derived in the $\Lambda$CDM-like
analysis is conservative.

%cosmological threads which influence on galaxy clusters with mass $M_{500}=2.5\times10^{14}\,h^{-1}\,M_{\odot}$

%\textbf{//////////////////////}\par

%According to figure~\ref{fig:33}, characteristic scales of these two effect are comparable even if gravity modification is rather small. Really, if we make $f(R)$ %gravity more significant and increase $\lambda$, the Compton wavelength increases too and influence of gravity modification to structure with scale 1-10 Mpc %becomes even stronger whereas characteristic scale of neutrino free-streaming remains the same. Conclusively, we find out that free-streaming damping and $f(R)$ %enhancement compete with each other on scales 1-10 Mpc for wide range of $\lambda$ .\par

 %Characteristic scale of galaxy clusters estimated as nearly 1-3 Mpc today. From figure~\ref{fig:33} in subsection~\ref{subsec:fluctLGC} we see that galaxy %clusters are indeed sensitive to these both counteractive forces in regions $z<2$ and $0.4<z<0.9$ where clusters are measured in case of near 1 eV neutrino mass. %Therefore, using of cluster mass function measurements to explore the net result of free-streaming erasement and $f(R)$ enhancement is allowable.\par

Recently a number of papers have been issued where the massive
neutrinos are exploited to suppress extra growth of matter density
perturbations at small scales occurring in various $f(R)$ models. For
example, while this work was finishing, the paper~\cite{B0last} with
the latest constraints on the parameter $\log_{10}B_{0}<-4.1$
(2$\sigma$) has appeared. Actually, this restriction has nothing to do
with our constraint because in that paper, first, no fourth sterile
neutrino is considered and, second, the BZ parametrization that is
formally similar to the case $n=0$ in our notation (actually, the
corresponding term in the action is even growing logarithmically with
$R$) is used. In the other interesting
article~\cite{f(R)_massivnu_last} it was found that the $f(R)$
model~\cite{Starmodel} increases the sum of the active neutrino masses
significantly that seems to be in conflict with our consideration. In
fact, the authors used another data set constraining the matter power
spectrum in contrast to galaxy clusters data set constraining
$\sigma_{8}$ used in our paper. The recent cluster mass function
measurements permit a rather high value of the sterile neutrino mass
without any modification of gravity due to the decrease in the value
of $\sigma_{8}$. The implementation of $f(R)$ gravity does not bring a
significant effect here because even higher mass of sterile neutrino
is forbidden by the combined CMB spectrum and BAO data set. At last, in
the work~\cite{f(R)_clusters} galaxy cluster measurements were used
with the mass function enhancement according to modified gravity
as compared to the $\Lambda$CDM consideration used here but once more,
for the standard number of neutrino species only. In this paper, the
Hu--Sawicky model~\cite{chameleonmech} is used which is very similar to
our model~\cite{Starmodel} and has the same behaviour at large $R$
with $2n$ denoted by $n$. Thus, our results for $n=2$ without the
sterile neutrino have to be compared with theirs for $n=4$. Earlier,
N-body calculations of the non-linear matter power spectrum, the halo
mass function and the halo bias with massive neutrinos were made in
\cite{Baldi2013} for the Hu--Sawicky $f(R)$ model in the case $n=1$
that corresponds to $n=0.5$ in our model. In all
papers~\cite{B0last,f(R)_massivnu_last,f(R)_clusters,Baldi2013} 3
standard neutrino species were assumed.

%However, it is worth noting that we reproduce neutrino mass constraint from this article in case of negligible modified gravity (high values of $B_{0}$) %according to

To understand which gravity law is more preferable by cosmological
data, we compare differences of logarithmic likelihoods $\log L$
calculated for different Universes with the same data set; for the
latter we chose the $ePlanck$+$BAO$+$LENS$+$H_{0}$+$CL$ set combination. Each difference $2\cdot\Delta\log L$ is distributed as $\chi^{2}$ with an effective number of degrees of freedom equal to the difference of
the numbers of fitting parameters in the two corresponding universes. The improvement of maximum likelihood for the $f(R)$ model ($\lambda$ for
the extra fitting parameter) with one massive
sterile and three massless active neutrinos as
compared to the $\Lambda$CDM model with one free massive and two massless
active neutrinos (normal hierarchy pattern) is $\Delta\log L=0.85$
which corresponds to $\chi^{2}=1.71$ for 1 degree of
freedom. Significance of such improvement is about 1.3$\sigma$.
Therefore, the Universe with one additional massive sterile neutrino
within $f(R)$ gravity is slightly more preferable than the
$\Lambda$CDM model with 3 active neutrinos (assuming normal
hierarchy). On the other hand, this would not be true if sterile species were not included into the modified gravity model. Indeed, implementation of $f(R)$ gravity in the case of only three
active neutrinos (with one massive) spoils the goodness-of-fit with respect to the $\Lambda$CDM model with the same neutrinos: $\Delta\log L=-3.42$ or $\chi^{2}=-6.83$ for 1 degree of freedom. The reason is that the CMB+BAO data combination constrains the mass of neutrino in case of 3 active ones more tightly in comparison with the model with 4 neutrino species (see discussion in Sec.~\ref{subsec:result}).

Moreover, $f(R)$ gravity significantly better describes the Universe
with the sterile neutrino of mass $\approx 1.5$\,eV introduced
for explanation of various anomalies in neutrino oscillation
experiments, as discussed in Introduction. We compare $f(R)$ gravity
with $\Lambda$CDM model for fixed sterile neutrino mass assuming
active neutrinos are massless. For sterile neutrino of $1$\,eV we
obtain a maximum likelihood ratio given by $\Delta\log L=-1.07$ which
corresponds to $\chi^{2}=-2.14$. It occurs because galaxy
cluster mass function data permit higher (though not too high) neutrino masses very
well without any gravity modifications (see Fig.~\ref{fig:66}). For
sterile neutrino of $1.5$\,eV, we find the improvement $\Delta\log L=9.52$
for the $f(R)$ gravity model with one additional free parameter $\lambda$
beyond GR. We see that although $f(R)$ implementation is not so
obvious for the $1$\,eV sterile neutrino, in the case of the $1.5$\,eV
mass (if found in ground experiments indeed) modified gravity improves the goodness-of-fit significantly. According to the Akaike Information Criteria (AIC)
\cite{AIC}, if $\chi^{2}$ improves by 2 or more with a new
additional free parameter, its incorporation is justified. In the case
of the $1.5$\,eV sterile neutrino, $f(R)$ implementation yields the improvement
$\Delta\chi^{2}=19.05$ for one additional free parameter.

\section{Conclusion}
\label{sec:concl}

In this work, we have reviewed $f(R)$ formalism generally and the
$f(R)$ cosmological model of the present DE~\cite{Starmodel} in particular. We have
used the First Iteration Approach (FIA) to describe the FLRW background
evolution in $f(R)$ gravity. Precision and range of
application of this method were studied in the case of
the~\cite{Starmodel} model.\par

We used CosmoMC package with the modified MGCAMB module to
  investigate the role of ${\cal O}(1)$ eV sterile neutrino in
  modified gravity using up-to-date cosmological data including low-z
  galaxy cluster mass function measurements. We do not find strong
degeneracy between the sterile neutrino mass and the parameter
$\lambda$ of the $f(R)$ gravity model used as suggested
before. Moreover, the $f(R)$ gravity effect on different parameter
constraints is not significant. Surprisingly, the existence of the sterile neutrino shifts the scalar spectral index $n_s$ to a value very close to unity irrespective of the law
of gravity used: GR or  $f(R)$ gravity. More importantly, modified gravity improves the maximum likelihood significantly for the fixed sterile neutrino mass $\approx
1.5$\,eV which is suggested by various anomalies in neutrino
oscillation experiments. Along with the fact that this modified
gravity does not spoil the goodness-of-fit for lower sterile neutrino
masses, $f(R)$ gravity remains more preferable in the description of the
Universe.\par

%\appendix
%\section{Some title}
%Please always give a title also for appendices.

\acknowledgments

The work was partially supported by the RFBR Grant No. 14-02-00894,
by the Scientific Program "Astronomy" of the Russian Academy of Sciences and by the Russian Government Program of Competitive Growth of Kazan Federal University. In this work the results of computations made with MVS-10P supercomputer of Joint Supercomputer Center of the Russian Academy of Sciences (JSCC RAS) were used.

%\paragraph{Note added.} This is also a good position for notes added
%after the paper has been written.

\end{document}